\documentclass[10pt,conference]{IEEEtran}
\IEEEoverridecommandlockouts
\usepackage{cite}
\usepackage{amsmath,amssymb,amsfonts}
\usepackage{algorithmic}
\usepackage{graphicx}
\usepackage{textcomp}
\usepackage{xcolor}
\usepackage{blindtext}
\usepackage{hyperref}
\usepackage{caption}
\usepackage{array}

\usepackage{textcomp}
\usepackage{hyperref}

\usepackage{comment}
\usepackage{xtab} % <-- Add this package
\usepackage{lipsum} % For dummy text
\def\BibTeX{{\rm B\kern-.05em{\sc i\kern-.025em b}\kern-.08em
    T\kern-.1667em\lower.7ex\hbox{E}\kern-.125emX}}
    
\begin{document}

\title{\fontsize{24pt}{28pt}\selectfont A Systematic Mapping Study on Risks and Vulnerabilities in Software Containers}
% Revealing practitioners' perspective on container security

\author{
    \IEEEauthorblockN{
        {1\textsuperscript{st}Maha Sroor}\IEEEauthorrefmark{1}, 
        {2\textsuperscript{nd} Teerath Das}\IEEEauthorrefmark{1},
        {3\textsuperscript{rd} Rahul Mohanani}\IEEEauthorrefmark{2},
        {4\textsuperscript{th} Tommi Mikkonen}\IEEEauthorrefmark{1}
    }
    \IEEEauthorblockA{
        \IEEEauthorrefmark{1}Faculty of Information Technology, University of Jyväskylä, Finland \\
        \IEEEauthorrefmark{2}M3S, University of Oulu, Finland \\
        \{maha.m.sroor, teerath.t.das, tommi.j.mikkonen\}@jyu.fi, rahul.mohanani@oulu.fi
    }
}

%\centering
%\author{\IEEEauthorblockN{First Author\IEEEauthorrefmark{1}, Second Author\IEEEauthorrefmark{2}, Third Author\IEEEauthorrefmark{3}, Fourth Author \IEEEauthorrefmark{4}}
%\IEEEauthorblockA{
%\textit{Institution}\\
 %Affiliation\\
%\{author1, author2, author3, author4\}@institution}

%\and

%\IEEEauthorblockN{Sandun Dasanayake}
%\IEEEauthorblockA{
%\textit{Faculty of ITEE}\\
%University of Oulu \\
%Finland \\
%sandun.dasanayake@oulu.fi}

%\and

\maketitle

\begin{abstract}

Software containers are widely adopted for developing and deploying software applications. Despite their popularity, major security concerns arise during container development and deployment. Software engineering (SE) research literature reveals a lack of reviewed, aggregated and organized knowledge of risks and vulnerabilities, security practices, and tools in container-based systems development and deployment. Therefore, we conducted a Systematic Mapping Study (SMS) based on 129 selected primary studies to explore and organize existing knowledge on security issues in software container systems. %Therefore, we conducted a Systematic Mapping Study (SMS) to comprehensively explore and organize the existing knowledge on the issues concerning the security of software container systems. % We identified 86,584 studies published between 2000 and 2024 across seven databases.A total of 129 relevant primary studies were selected after applying the selection criteria. 
Data from the primary studies enabled us to identify critical risks and vulnerabilities across the container life-cycle and categorize them using a novel taxonomy. Additionally, the findings highlight the causes and implications and provide a list of mitigation techniques to overcome these risks and vulnerabilities. Furthermore, we provide an aggregation of security practices and tools that can help support and improve the overall security of container systems. This study offers critical insights into the current landscape of security issues within software container systems. Our analysis highlights the need for future SE research to focus on security enhancement practices that strengthen container systems and develop effective mitigation strategies to comprehensively address existing risks and vulnerabilities.

\end{abstract}

\section{Introduction}
\label{Introduction}

Software containers, often referred to simply as containers, have emerged as a popular solution for deploying software applications \cite{sroor2022leverage}. Containers are widely recognized for enhancing software agility, accelerating deployment cycles, and supporting the development of scalable microservice architectures \cite{hasselbring2017microservice}. As a result, they have emerged as a significant advancement in software engineering (SE), particularly following the limitations of traditional monolithic application architectures \cite{bhardwaj2021virtualization}. Consequently, containers are strongly advocated in modern contemporary software application development practices \cite{crnkovic2001component} \cite{reifer2002good}.

However, despite the multiple advantages containers offer to software development, numerous concerns regarding their security persist. Indeed, any significant impact on security stands out as a primary barrier to widespread container adoption \cite{combe2016docker}. Hence, an urgent need exists to investigate container security and explore the contributors behind risks and vulnerabilities within the container environment \cite{sultan2019container}. In the context of this research, ``\textit{risks}'' denote potential attack vectors \cite{taherdoost2021review},
while ``\textit{vulnerabilities}'' refer to the systems' weaknesses that are susceptible to exploitation \cite{aslan2023comprehensive}. Identifying and mitigating these risks and vulnerabilities are critical for strengthening container integrity, improving trust, and fostering its adoption.

Underlying risks and vulnerabilities in software container systems are critical because containers are not a stand-alone technology by themselves. They rely on a host machine to communicate with other entities, such as the network and orchestration platforms \cite{bernstein2014containers}. Examining these risks and vulnerabilities, along with their root causes, impacts and mitigation techniques, is essential for improving the overall security of the software system. This understanding will inform security decisions in container systems \cite{vsijan2024modeling}, contributing to a safer and more secure software development environment.

This study presents the findings from a Systematic Mapping Study (SMS) on state-of-the-art SE research literature concerning security issues in software containers. It specifically focuses on identifying potential risks and vulnerabilities that could impact the security of containerized environments. Furthermore, our analysis provides an in-depth examination of the underlying causes of these risks and vulnerabilities, their effects and mitigation techniques. Additionally, it explores security practices and tools that effectively enhance overall container security. Here, ``\textit{security practices}'' refers to the collective processes and techniques that aim to improve container systems\textbf{'} security \cite{balzacq2010security}. Whereas, ``\textit{tools}'' are the artifacts that can help to maintain and monitor container systems within their life-cycle \cite{reiss1996software}. Container \textit{``life-cycle''}, in the context of this study, refers to the distinct but interrelated phases of developing, deploying, and running container systems. It consists of choosing the container image, configuring the host machine, implementing the container and its applications, setting the network and orchestration platform, and container runtime (based on \cite{vs2023container}).

%To the best of our knowledge, no comprehensive and systematic literature review has specifically focused on risks and vulnerabilities in software container systems, along with their mitigation techniques, security tools, and best practices. In addition, only a 
To the best of our knowledge, few existing studies have attempted to categorize risks and vulnerabilities related to the development and deployment of software containers. This study introduces a novel and comprehensive taxonomy that spans all phases of the container life-cycle, offering a more extensive classification compared to previous works on container security risks and vulnerabilities. For example, \cite{ugale2023container} and \cite{souppaya2017application} examined risks and vulnerabilities in specific phases of the container life cycle. Additionally, \cite{vs2023container} focused on the Docker and Podman life-cycle only without investigating container orchestration risks and vulnerabilities. Moreover, \cite{wong2023security} employs the STRIDE framework to identify risks in the container environment and propose mitigation strategies. The STRIDE framework identifies security threats in software system architectures. It focuses on six categories of threats only (spoofing, tampering, repudiation, information disclosure, and denial of service).

Given the popularity of containers, exploring their security aspects is essential for developing software applications and ensuring their quality and integrity. Consequently, the primary objective of this research is to provide a comprehensive summary of the current state-of-the-art research on software container risks and vulnerabilities by focusing on their identification and understanding the root causes, effects, and mitigation. In addition to exploring the tools and practices used to enhance container security.

% To maintain the scope of this secondary review of relevant SE literature, this study does not aim to collect and analyze any empirical data to support or test the effectiveness of the techniques mentioned in Section \ref{Results}. 

%Consequently, the primary objective of this research is to provide a comprehensive summary of the current state-of-the-art research on software container risks and vulnerabilities by focusing on their identification, mitigation, and exploring the tools and practices used to enhance container security. To maintain the scope of this secondary review of relevant SE literature, this study does not aim to collect and analyze any empirical data to support or test the effectiveness of the techniques mentioned in Section \ref{Results}. 

The main research question guiding this study is: \textit{What is the state-of-the-art in the existing literature on risks and vulnerabilities in software container systems?}

Our study makes the following novel contributions to the knowledge of security concerns in software containers:
\begin{enumerate}
%\item we could not identify any systematic mapping study discussing container risks and vulnerabilities in container systems in literature; our team was the first to do similar work.

\item Proposes a novel taxonomy for categorizing the risks and vulnerabilities throughout the container development life-cycle.
%Proposes a novel taxonomy for categorizing the risks and vulnerabilities throughout the container development life-cycle. This taxonomy will facilitate the identification of the phase in which the specific risks or vulnerabilities originate. 

\item Systematically identifies potential risks and vulnerabilities in software containers, along with their causes, implications and mitigation techniques.
%Systematically identifies potential risks and vulnerabilities in software containers, along with their causes, implications and mitigation techniques. Identifying and considering the causes and implications of these risks and vulnerabilities would guide the appropriate strategy and techniques for their mitigation. 

\item Explores and highlights various container security practices that protect containers from (potential) risks and vulnerabilities.
%. This will support to protect container systems from (potential) risks and vulnerabilities.
   % summarises the security practices that help to improve the container systems' security and address the mitigation techniques to avoid container risks and vulnerabilities; and
\item Provides a list of container security tools with an examination of their functionalities, helping to maintain security within the container life-cycle. 
   
\end{enumerate} 

The study is structured as follows: Section \ref{background} provides a short overview of containers and container security. Section \ref{studydesign} gives in-depth information on the guiding research questions and the research protocol. Section \ref{Results} reports findings from the literature on container risks and vulnerabilities, security practices used to improve security, and tools used to maintain security in a container environment. Section \ref{Discussion} provides our interpretation of the results and a future research roadmap. Section \ref{Conclusion} highlights the main study contributions. 

\section{Background}
\label{background}

%This section introduces the background and builds the scope of the subsequent research. 

%including brief information on software containers, container security, and motivation for this research. In the following, each of these topics is discussed separately.

%\subsection{Software Containers}

Containers have become essential in software deployment, transforming how software is developed and executed \cite{kithulwatta2021adoption}. 
They encapsulate packages of applications and their dependencies, providing an isolated system for running diverse types of software \cite{bentaleb2022containerization}. Containers support building, testing, and deploying applications across various environments, such as development, staging, and production \cite{anderson2015docker}.

A container is a lightweight and portable software unit that packages application code, dependencies, configuration files, and the runtime environment, enabling it to execute in an isolated environment \cite{docker2024}. %, \cite{hamzaoui2024topical}, \cite{li2025lightweight}. 
The typical structure of containers involves multiple applications running on top of a shared container engine. The container engine operates above a host operating system and a physical server \cite{bhardwaj2021virtualization}. Several container technologies, such as LXC \footnote{https://linuxcontainers.org/lxc/}, openvz \footnote{https://openvz.org/}, Apptainer \footnote{https://apptainer.org/} (formerly known as Singularity), and udocker \footnote{https://github.com/indigo-dc/udocker} are only supported by Linux, while others like Docker \footnote{https://www.docker.com/} supports multiple operating systems, including Windows, Linux, and macOS \cite{malhotra2024systematic}.

%Containers operate seamlessly by connecting to the host operating system and reserving file systems, CPU, and memory to create an isolated system for running applications \cite{maenhaut2020resource}, \cite{patel2021interval}.  Additionally, the software container features a virtual private network interface that enables the execution of private network connections \cite{hoenisch2015four}.

Software containers offer numerous advantages for software development and its deployment. For example, they reduce resource consumption by sharing resources with the host operating system. Additionally, they save time due to lower overhead compared to virtual machines. Furthermore, containers enhance accessibility by eliminating bugs related to software transactions in the runtime environment. Moreover, they are lightweight and support DevOps, facilitating quick emulation to a production-ready distributed system \cite{paraiso2016model}, \cite{dua2014virtualization}, \cite{han2025dmscts}.

%Despite the countless advantages of containers, there are critical concerns regarding containerized application security \cite{jolak2022conserve}. Addressing these concerns is crucial to improving their adoption and encouraging container migration \cite{sultan2019container}. 
%container security remains a major concern, as breaches can impact application performance and availability.
%\subsection{Container Security }
%\label{Relatedwork}

Despite the countless advantages of containers, container security is considered a major concern \cite{sultan2019container}. Numerous studies have discussed container security from various perspectives \cite{martin2018docker, kaur2021analysis, wenhao2020vulnerability, shamim2020xi, rahman2023security}. Most existing studies focus on specific phases or components of the container environment, as we elaborate further in this section.

%because compromising container security can seriously affect application performance and availability \cite{sultan2019container}. Numerous studies have discussed container security from various perspectives \cite{martin2018docker, kaur2021analysis, wenhao2020vulnerability, shamim2020xi, rahman2023security}. The studies studied specific parts of containers' risks and vulnerabilities during their life-cycle.

%\cite{martin2018docker} \cite{kaur2021analysis} \cite{wenhao2020vulnerability} \cite{shamim2020xi} \cite{rahman2023security}

Martin et al. \cite{martin2018docker} identify five categories of container vulnerabilities. The first category discusses how container misconfiguration in production environments can compromise the host and lead to a denial of service. The second category addresses vulnerabilities in image distribution, repositories, and verification that may result in unauthorized access to sensitive data. The third category focuses on vulnerabilities within the image that potentially allow control over the container and its data. The fourth category examines vulnerabilities in libraries and lib-containers, highlighting risks to the Docker host. Finally, the fifth category addresses Linux kernel vulnerabilities that could grant full root privileges to containers. The article comprehensively analyzes Docker environment security, covering images, hosts, and container vulnerabilities.

Kaur et al. \cite{kaur2021analysis} compare the results of four vulnerability scanners on container images. While the scanning results varied significantly, they all revealed the presence of numerous vulnerabilities in the images. Some of these vulnerabilities are critical and pose a threat to container systems. The majority of vulnerabilities stem from outdated packages. The article proposes two effective solutions to mitigate these vulnerabilities: regular updates of software packages and removal of unused software packages. These solutions reduce the number of vulnerabilities significantly. Overall, the article provides an insightful analysis of image vulnerabilities in container systems.  Similar results were presented in \cite{shu2017study}, \cite{javed2101understanding}, indicating that the majority of vulnerabilities stem from outdated packages.

Wenhao et al. \cite{wenhao2020vulnerability} analyze Docker's research on security, focusing on vulnerability analysis. From the author's perspective, vulnerability analysis centres around isolation and resource sharing. The article delves into isolation in file systems, communication, host resources, and the network. Additionally, the author dedicates a section to discussing kernel security and its inherited options that can enhance security in container systems when properly configured.
The author discusses the role of Linux functions, AppArmor, capability mechanisms, Seccomp, access control, integrity protection, and log auditing. This article highlights isolation issues and emphasizes the crucial role of the host system in protecting overall container security, similar to the results from \cite{bazm2019isolation}. 

Jarkas et al. \cite{jarkas2025container} investigated the most frequently occurring vulnerabilities, such as \textit{container escape}, \textit{image tampering},\textit{ runtime vulnerabilities}, \textit{privilege escalation}, and \textit{ namespace bypass}, based on existing research in academia and industry. The article identifies 47 distinct categories of container vulnerabilities across 11 attack vectors based on attack mechanism. The article's findings are highly focused on the vulnerabilities related to multi-tenant cloud settings. Moreover, it provides actionable insights to help practitioners improve container security and mitigate possible exploits.

Shamim et al. \cite{shamim2020xi} attempted to systemize knowledge on Kubernetes security practices that are used to improve its security, where a qualitative analysis was applied to 104 internet artifacts, identifying 11 security practices. The most effective security practices discussed in the article included implementing role-based access control (RBAC) authorization to reduce privileges, implementing security policies for networks and pods, and updating Kubernetes with security patches.

Rahman et al. \cite{rahman2023security}  aimed to secure the Kubernetes cluster by identifying security misconfigurations. In the study, 92 open-source software repositories were examined to identify 11 categories of security misconfiguration. The authors also investigated the frequency of these misconfigurations and the system objects affected by them. The findings from the article are similar to findings in \cite{mustyala2021advanced}, which ensure the role of proper configuration management in maintaining the security of Kubernetes clusters.

This study aims to develop a comprehensive understanding of container risks, vulnerabilities, security practices used to improve security, mitigation techniques, and security tools used to maintain security. Furthermore, we intend to explore the broader container environment, examining its role in introducing container risks. We will also identify the primary causes and implications of these risks and vulnerabilities. This enhanced understanding of multiple aspects of risks and vulnerabilities is vital for fostering a safer software development process and meeting the growing demand for software applications \cite{donca2022method}, \cite{perumal2021enhancing}.

%\subsection{Research Motivation}

%Containers provide a solution for developing and deploying applications by offering a virtualized and isolated software environment for running applications and their dependencies \cite{koskinen2019containers}. Given the popularity of containers, exploring their security aspects is essential for developing software applications and ensuring their quality and integrity. A literature review reveals numerous studies discussing container security, but many of these discussions are rather general and do not address specific security issues. Additionally, while some studies focus on the security issues inherent to containers, only a few discuss the container environment's significant influence in triggering these issues.

\section{Study design}
\label{studydesign}

%This section describes the methodology used to achieve the objective of our study. It describes the entire research process used to collect and synthesize the primary studies. It details the data collection, synthesis, and analysis methods to ensure transparency and reproducibility.

\subsection{Research Questions}

%Based on the primary objective, the main research question guiding this study is---\textit{What is the state of the art in identifying risks and vulnerabilities in software container systems, along with their causes, effects, mitigation techniques, security practices, and tools?}

Exploring container security is essential for improving container adoption in software deployment. Therefore,  the main research question guiding this study is---\textit{What is the-state-of-the-art in the existing literature on risks and vulnerabilities in software container systems?}

Considering the comprehensive aspects addressed in this research, the main research question is subdivided into multiple sub-research questions, where each sub-research question addresses one aspect of container security in detail. Table \ref{tab:Research Questions and Reasoning} presents the sub-research questions and their corresponding reasoning based on the research objective (as specified in Section \ref{Introduction}.

\begin{table*} [htp]
 \centering
  \caption{Sub-Research Questions}
 \label{tab:Research Questions and Reasoning}
\setlength{\tabcolsep}{2pt}
\begin{tabular}{|p{160pt}|p{330pt}|}
\hline
\textbf{Sub-Research Questions} & \textbf{Justification} \\
\hline
\textbf{RQ 1:} What are the risks and vulnerabilities associated with software container systems? & %This question seeks to identify and articulate the risks and vulnerabilities inherent in the container life cycle that include an image, host, intra-container, network and orchestration, and run time. The question also details the causes and effects of each.
  Seeks to identify and categorize the risks and vulnerabilities inherent across the container life-cycle, detailing the causes, effects, and mitigation techniques of each risk and vulnerability.
      \\
  \textbf{RQ 2:} What are the security practices to avoid risks and vulnerabilities in container systems? & Explores various security practices that can be implemented to safeguard container systems against potential risks and vulnerabilities. \\
    \textbf{RQ 3:}What are the tools used to improve container security? &Investigates the range of tools utilized to enhance container systems’ security.\\
   \textbf{RQ 4:} When was the article published? & Traces the publication trend concerning container security, highlighting how the topic has evolved.\\
  \textbf{RQ 5:} Where were the articles published? & Identifies the venues of these primary studies, providing insights into their quality level, and ensuring they are peer-reviewed.\\
  \textbf{RQ 6:} What are the most frequently used research methodologies? & Highlights the diversity in research methodologies employed in container security.\\     
  \hline

\end{tabular}
%\label{tab1}
\end{table*}

\subsection{Research Method}
\label{Research Method}

We chose the Systematic Mapping Study (SMS) approach to comprehensively understand container security's current state-of-the-art by focusing on the risks, vulnerabilities, and techniques for their mitigation. This approach was adopted to provide a high-level perspective and comprehensively cover the unexplored topic of container security. An SMS approach would further help to systematically aggregate and categorize existing knowledge, laying the foundation for analyzing container risks and vulnerabilities. 

To ensure the accuracy of our research process, we followed the process as prescribed in \cite{kitchenham2010systematic} and \cite{petersen2015guidelines}. Additionally, we considered ACM SIGSOFT empirical standards for mapping studies \cite{ralph2020acm} to ensure compliance with the requirements for systematic studies.

\subsection{Research Protocol}
The entire author team planned and agreed on the research protocol. It consists of the following main steps.

\begin{enumerate}
    \item Database selection and search string formulation (refer to Section \ref{Literature search}).
    \item Removing duplicate articles (refer Section \ref{de-duplication}).
    \item Automated and manual screening of the retrieved articles (refer to Section \ref{Screening}).
    \item Apply recursive backward snowballing (refer Section \ref{Ensuring Sample Adequacy}).
    %\item Assess the quality of the primary studies \ref{Quality Assessmen}.
     \item Extract the relevant data to address the research questions (refer to Section \ref{Data Extraction}).
    \item Analyse the extracted data (refer Section\ref{data Analysis}).
    
\end{enumerate}

\subsubsection{Database selection and search string formulation}
\label{Literature search}

As suggested in \cite{dyba2007applying}, we used the commonly searched databases for SE studies, including IEEE Explore, ACM Digital Library, Scopus, Web of Science, and Science Direct, enduring a thorough exploration of the relevant articles. Our search process involved three separate pilot runs to have a common understanding of the search process and ensure complete coverage of the search as much as possible. The pilot and search process is described as follows.

    \noindent (1) \textit{Pilot 1:} We employed the research term \textit{``software AND container* AND risk* OR vulnerabilit* OR security''} on the metadata. The database results yielded 1,347,088 primary studies, leading to unmanageable datasets. After analyzing a sample of randomly selected 100 articles across all databases, it was observed that the majority belonged to other domains and research disciplines. To refine our search, we decided to remove the term ``security'' from the search string, given its broad scope.

\noindent (2) \textit{Pilot 2:} Based on the earlier pilot, we decided to revise the research string as \textit{``software AND container* AND risk* OR vulnerabilit*''}. We also used the available database filters for more refined results. After that, we compared the search results against an initial reading list of relevant studies and found that only 60\% of the articles in the initial reading list were covered. This analysis highlighted the necessity of expanding our search scope to include Springer and Wiley databases. Furthermore, we acknowledged the significance of searching the full text rather than restricting the search to metadata alone.
\begin{comment}
    
\begin{table*} [htp]
   \caption{Search Pilot 2 }\label{table:filters_results_trial2}
 \centering
\setlength{\tabcolsep}{4pt}
\begin{tabular}{|p{65pt}|p{160pt}|p{160pt}|p{65pt}|}

\hline
Database & String  & Filters & Results \\
\hline
IEEE Xplore & software AND
 container* AND risk* OR vulnerabilit* & Search Metadata, search from 2000, subject area: cloud  computing, network security& 94  
\\

Scopus & software AND container* AND risk* OR vulnerabilit* & Subject area: computer science, search from 2000 & 374 
\\

ACM & software AND container* AND risk* OR vulnerabilit* & Subject area: computer science, search from 2000 & 510 \\

Web of Science & software AND container* AND risk* OR vulnerabilit*  & Subject area: computer science, search from 2000 & 1923\\

Science Direct & software AND container* AND risk* OR vulnerabilit* & Search starts in 2000, Subject area: computer sciences & 998 \\ 
%\\
\hline
Total &  &  & 3899\\        
\hline  

\end{tabular}
%\label{tab1}
\end{table*}
\end{comment}

\noindent (3) \textit{Pilot 3:} Building upon our previous insights, we conducted a full-text search to broaden our scope and ensure the inclusion of a more comprehensive selection of articles using the query ``\textit{software AND container* AND (risk* OR vulnerabilit*)}''. The articles were retrieved in the period from 2000 to 2024. Additionally, we employed database-specific filters to refine our search parameters, as detailed in Table \ref{table:filters_results_trial3}.

\begin{table} [htp]
 \caption{Search Pilot 3 }
\label{table:filters_results_trial3}
 \centering
\setlength{\tabcolsep}{4pt}
\begin{tabular}{|p{45pt}|p{150pt}|p{20pt}|}
\hline
\textbf{Database}& \textbf{Filters} & \textbf{Results} \\
\hline
IEEE Xplore & Searched Full text and metadata,
year: 2000 to 2024, publication source: conference, journal articles, magazine, early access articles, standards &  11923 \\

Scopus &Searched Full text, limit search to year:2000 to 2024, subject area: computer science and engineering, source type: journal and conference proceeding, article type: article, conference paper, review, conference review, editorial, and data paper, language: English &  3933 
 \\

ACM & Searched full text, year:2000 to 2024, content type: research article, short paper, journals & 54,292 \\

Web of Science & Searched all fields /No Filters used & 133 \\

Science Direct & Year:2000 to 2024, article type:
research articles, correspondence, data article, editorial, field: computer sciences, Engineering & 6557 \\

Wiley & Searched anywhere in context, year:2000 to 2024, document type: journal reference work, subject: computer sciences & 7572  \\

Springer & Year: 2000- 2024, area: computer science, articles type: conference paper in computer science, conference proceeding in computer sciences, conference paper in engineering, conference proceeding in engineering, articles in engineering & 2263  \\
\hline
Total &  & 86673  \\
\hline

\end{tabular}
%\label{tab1}
\end{table}

Upon analyzing the results, we achieved 100\% coverage against an initial list of relevant studies. However, a notable issue arose: a substantial portion of our primary studies originated from outside the SE domain. After deliberation within the author's team, we opted to proceed with the current results to ensure the inclusion of all potential primary studies. We retrieved a total of 86673 articles from our search.

\subsubsection{De-duplication}
\label{de-duplication}

We utilized a spreadsheet to store, organize, and manage the retrieved articles. By employing the available automated function within the spreadsheet, we conducted a deduplication process on the titles of the primary studies, successfully removing 23,113 duplicate articles. Following this de-duplication procedure, we were left with 63,560 articles poised for the screening process to select the relevant primary studies.

\subsubsection{Screening} \label{Screening}

\noindent \textbf{Automated Screening}.
Recognizing that manually screening (i.e., employing the inclusion and exclusion criteria on) 63,560 articles would be time-consuming, we decided to expedite the process by using an automated function within the spreadsheet. After a quick analysis of randomly selected articles, we found that articles not featuring the terms---``container'' or ``docker''---at least once in their titles or abstracts were irrelevant to our research.

To streamline the exclusion of irrelevant studies, we formulated a query to automatically exclude articles lacking the terms ``docker'' or ``container'' in their abstracts or titles. The query utilized was:
\textit{``\textbf{=SUMPRODUCT(--ISNUMBER(SEARCH({container'';docker''};E4)))$>$0}''}.
The automated exclusion process successfully removed 58,661 articles from consideration, leaving us with 4,899 articles to undergo the manual screening process. \\

\noindent \textbf{Manual Screening}. Here, we developed and applied specific inclusion and exclusion criteria to manually filter the relevant studies. The inclusion and exclusion criteria are as follows.
\medskip
\noindent \textit{Inclusion criteria}: 
\begin{enumerate}
    \item Peer-reviewed articles published in scientific journals, conferences, and workshops.
    \item Articles belonging to the SE domain.
    \item Security-focused studies on software containers.
\end{enumerate}
\medskip
\noindent \textit{Exclusion criteria}: 
\begin{enumerate}
    \item Articles not written in English.
    \item Books, secondary and tertiary studies, magazine articles, reviews, posters, master's and doctoral thesis.
\end{enumerate}

We screened the studies initially based on their titles, followed by the metadata, introduction \& conclusion, or by reading the entire text if at all was deemed necessary.

To achieve a common and consistent understanding of the topic and the screening process, the first and second authors conducted a pilot process to assess the agreement levels. 42 randomly selected articles were split between the two authors, who independently applied the inclusion and exclusion criteria. This produced Cohen's kappa score of 0.92, indicating high agreement between the first and second authors \cite{cohen1968weighted}. The third and fourth authors actively monitored and reviewed the entire pilot process and resolved the disagreements.

After manually applying the process, we were left with 161 articles. Additionally, six articles were excluded due to limited access to their content, and another four short/early-ideas articles were excluded. At this stage, we had a total of 151 included articles.

\subsubsection{Recursive Backward Snowballing}
\label{Ensuring Sample Adequacy}

The reference list of the 151 articles was assessed to find any relevant articles that were not included in our primary study list. Backward snowballing was chosen for its ability to target foundational and highly impactful studies in container security, ensuring a focused scope and relevance to container risks and vulnerabilities. After using the manual screening process (as described earlier), we included 13 new articles. We subsequently examined the references of these newly included articles but found no relevant articles. At the end of this process, we included a total of 164 primary studies.  

\subsubsection{Data Extraction }
\label{Data Extraction}
We utilized the qualitative data analysis software Atlas.ti\footnote{https://atlasti.com/}, known for its advanced coding capabilities that facilitate the organization, analysis, and visualization of qualitative data. A data extraction pilot run was conducted on ten randomly selected primary studies, and the results were reviewed and confirmed by the second and third authors. 

An \textit{a-priori} coding scheme (see \cite{cruzes2011recommended}) was applied to segments of primary studies based on the research questions. Table \ref{tab: extracted Data Items} presents the data items, their definitions in the context of this study, and their relevance to the research questions. The entire author team thoroughly reviewed the coding schema.

\begin{figure}
    \centering
    \includegraphics[width=1\linewidth]{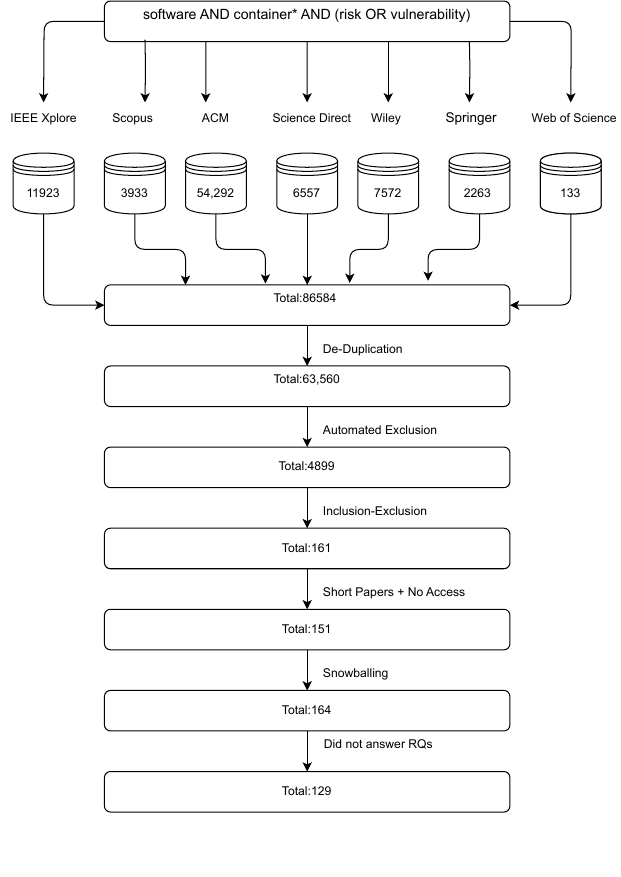}
   \caption{Primary Studies Collection and Synthesizing}
    \label{fig:primary studies}
\end{figure}

While extracting the relevant data, it became clearer that a few of the included studies did not contain information that answered our research questions. Therefore, we excluded 35 studies that did not discuss any of the sought-after information in the context of this research---e.g., container risks and vulnerabilities, mitigation techniques, or security practices. This resulted in a total of 129 primary studies. Fig. \ref{fig:primary studies} summarizes the entire primary study selection process. The final list of 129 primary studies can be accessed here: \url{https://zenodo.org/records/15813355}.

\begin{table*} [htp]
\caption{Extracted Data Elements}
\label{tab: extracted Data Items}
 \centering
\setlength{\tabcolsep}{4pt}
\begin{tabular}{|p{100pt}|p{320pt}|p{30pt}|}\hline
\textbf{Data Item} &\textbf{ Explanation}  & \textbf{RQ\# } \\
\hline
Vulnerability & System weaknesses that could be exploited to compromise the system & RQ1 \\

Risk & Potential attacks to the container system & RQ1\\

Mitigation techniques & Specific actions done to reduce the likelihood of a risk or a vulnerability & RQ1 \\

 Security practices & Policies that should be taken into action to improve the overall security of container systems  & RQ2\\
 
Security tool & Software application or program designed to perform specific tasks to improve container systems security & RQ3 \\

Year of publication & Year when the article was published  & RQ4\\
Venue of publication & Outlet where the primary study was published & RQ5\\

Research Method & Research methodology employed in articles & RQ6 \\
\hline
\end{tabular}
%\label{tab1}
\end{table*}

%\begin{figure*}
 %   \centering
  %  \includegraphics[width=0.5\linewidth]{Porotocol.png}
  %\caption{Primary Studies Collection and Synthesizing Process}
   % \label{fig:primary studies}
%\end{figure*}

\subsubsection{Data Analysis}
\label{data Analysis}
We organized the extracted data on risks, vulnerabilities and their mitigation techniques, and security tools according to a new taxonomy aligned with the container life-cycle, featuring five categories, namely (\textit{image}, \textit{host}, \textit{intra-container}, \textit{networking and orchestration}, and \textit{runtime}). More details on each category are provided across Section \ref{Results}.

%The first category is the \textit{image}, encompassing the inactive software package stored on public servers. The second category is the \textit{host}, covering the host infrastructure and its connection to the container. The third category is the \textit{intra-container}, involving the state of the software package in the running state and its interaction with the system in addition to the internal container interactions. The fourth category is \textit{networking and orchestration}, addressing container communication with the outside world and the management of container operations in the cloud. The fifth category is \textit{runtime}, representing the fully configured container system capable of providing the required service. More details on each category are provided across Section \ref{Results}.

\section{Findings}
\label{Results}

\subsection{Risks and vulnerabilities in container systems}
\label{RQ: What are container security risks and vulnerabilities?}

This section (addressing RQ1) seeks to explore container systems' myriad risks and vulnerabilities. Also, it aims to address the causes, effects, and mitigation techniques of each risk and vulnerability. We found 66 different risks and vulnerabilities organized into six categories based on the phases of the container life-cycle. %---\textit{image, host, intra-container, network and orchestration, and runtime}.
The last category contains all other risks and vulnerabilities that do not belong to any of the previous categories. 

\begin{figure}
   \centering
    \includegraphics[width=1\linewidth]{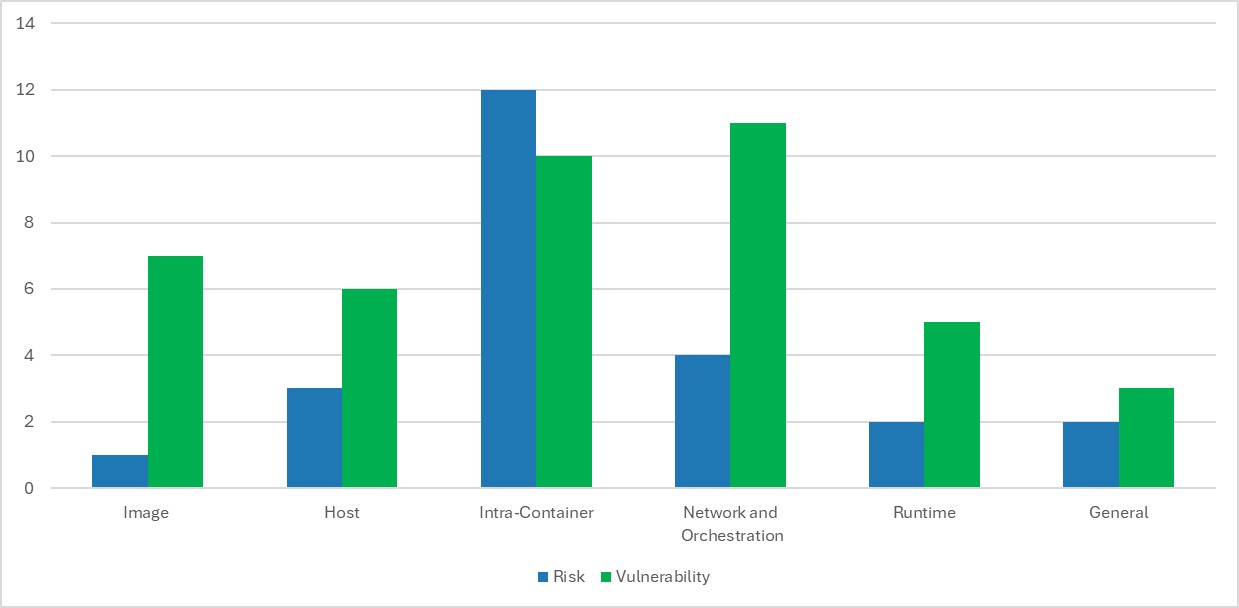}
    \caption{Number of Risks and Vulnerabilities in Container Life-Cycle Phases}
   \label{fig:Number of Risks and Vulnerabilities in Container Life-Cycle Phases}
\end{figure}

Fig. \ref{fig:Number of Risks and Vulnerabilities in Container Life-Cycle Phases} presents a summary of the distribution of risks and vulnerabilities across all phases of the container life cycle. The graph highlights a clear disparity in research focus, with the `image,' `host,' and `runtime' phases receiving comparatively less attention regarding investigations concerning security issues. The subsequent sections describe each category and present the potential risks and vulnerabilities that can arise in each phase. Table \ref{Primary studies discussed risk and vulnerabilities} lists the primary studies that discussed risks and vulnerabilities in container systems. More detailed information about the risks, vulnerabilities, causes, effects, and mitigation techniques in container systems is available at:  \url{https://zenodo.org/records/15813355}

%More detailed information on the risks, vulnerabilities, causes, impacts, and mitigation techniques across the container life cycle---along with the primary studies that discuss these aspects--- is available at the provided link \url{https://zenodo.org/records/15813355}.

%\begin{comment}

\begin{table} [htp]
 \caption{Primary studies}
\label{Primary studies discussed risk and vulnerabilities}
 \centering
\setlength{\tabcolsep}{2pt}
\begin{tabular}{|p{45pt}|p{170pt}|}
\hline
\textbf{Phase}& \textbf{Primary studies}\\
\hline

Image & [P3], [P4], [P5], [P7], [P10], [P12], [P24], [P32], [P51], [P55], [P65], [P70], [P82], [P95], [P97], [P98], [P99], [P107], [P108], [P118], [P121] \\

Host & [P3], [P10], [P11], [P18], [P23], [P32], [P36], [P40], [P57], [P65], [P84], [P88], [P95], [P103], [P112], [P114], [P117], [P127]  \\

Intra-Container & [P1], [P4], [P5], [P7], [P9], [P11], [P14], [P16], [P18], [P22], [P23], [P27], [P29], [P32], [P33], [P38], [P39], [P42], [P47], [P48], [P49], [P50], [P53], [P54], [P61], [P63], [P64], [P66], [P71], [P76], [P77], [P79], [P94], [P97], [P101], [P106], [P116], [P119], [P120], [P122], [P123], [P129] \\

Network and Orchestration & [P1], [P11], [P19], [P23], [P30], [P41], [P50], [P55], [P59], [P68], [P112], [P115], [P119], [P123] \\

Runtime & [P3], [P7], [P9], [P20], [P23], [P30], [P62], [P78], [P80], [P96], [P128] \\

General & [P3], [P41], [P84], [P85], [P106], [P119] \\

\hline

\end{tabular}
\end{table}

%\end{comment}

\subsubsection{Image}

A container image is a lightweight and executable software package essential for running an application and represents an immutable form of containers \cite{xu2018mining}. Such images can be obtained from cloud repositories such as Docker Hub \footnote{https://hub.docker.com/} and Amazon ECR Public Gallery \footnote{https://gallery.ecr.aws}.%Table \ref{Container Image  Security Vulnerabilities and Risks} presents a list of eight security risks and vulnerabilities associated with container \textit{images}.
Our findings identified eight security risks and vulnerabilities associated with container \textit{images}. These security concerns can be categorized into three groups:
\begin{itemize}
    \item \textit{Configuration flaws}: it refers to insecure settings in container images. It includes visual data authentication in container images, outdated node package manager (npm) packages, and unauthorized modification of the container image layer.
    \item \textit{Malicious image sources}:  refers to a container image that is designed to cause harm to the container system. It includes using untrusted image sources, image poisoning, running high-risk base images, and unused image functions.
    \item \textit{Exploitable repository architecture}: refers to vulnerabilities in the repository structure, making it vulnerable to attacks. It includes Docker Hub's architecture.
\end{itemize}

\subsubsection{Host}

The container host is the infrastructure that runs the container in the test or production environment. It supports containers with the necessary OS, memory, libraries, and network configurations required for its operation \cite{bernstein2014containers}. Our findings identified nine risks and vulnerabilities associated with the container host. Based on their causes, we present the risks and vulnerabilities, grouped into three major groups, as follows:

\begin{itemize}
    \item \textit{Host configuration risks and vulnerabilities}: This includes Kernel-Level exploits and an insecure host OS network stack.
    
    \item \textit{Host resource isolation risks and vulnerabilities}: This include insufficient host hardening.
    
    \item \textit{Escalated access permissions risks and vulnerabilities}: This includes executable rights in temporary /tmp Directory, unprotected grand unified boot-loader (GRUB), unprotected shared resources, binary attacks, cipher block chaining (CBC) attacks, and Linux attacks. 
% Mistake unprotected shared resources    
\end{itemize}

\subsubsection{Intra-Container}

Containers are a form of lightweight virtualization that enables the isolation and packaging of applications with their dependencies into a container. It does not work independently. It relies on a hosting machine to utilize its resources \cite{paraiso2016model}. We identified a total of twenty-two intra-container risks and vulnerabilities and grouped them based on their inherent causes, as follows:

%Due to the large number of intra-container risks and vulnerabilities, we report our findings across two separate tables---for risks and vulnerabilities---to enhance readability.

%Table \ref{tab:Risks and Vulnerabilities in `Intra-Container} present the list of intra-container vulnerabilities and risks, respectively. 

%Table \ref{tab:container vulnerabilities} addresses the ten container vulnerabilities according to the causes of the risks and vulnerabilities 
\noindent 
 \begin{itemize}
     \item \textit{Configuration vulnerabilities}: These include docker daemon socket misconfiguration, covert channels, root privilege escalation, namespace exploitation, and notary server verification bypass.
     \item \textit{Vulnerable container architecture}: refers to weaknesses in container structure. It includes central resource manager quality of service (QOS) blindness, and running a container on a virtual machine.
     \item \textit{Vulnerabilities due to unclear policies}: It includes a temporary file (TF )smell and an outdated seccomp profile.
    \item \textit{Unauthorized access risks}: they are the risks that happen because of unauthorized access. It includes the host gaining access to the guest containers, unauthorized access to containers, and container escape attacks.
    \item \textit{Risks due to intra-container attacks}: They are the attacks that target internal entities inside the container. They include unauthorised access to intra-container ``Secrets'', attacks on honeypots, and race attacks.
     \item \textit{Risks due to multifaceted attacks} They include distributed side-channel attacks (DSCA), surface attacks on containers, running the container on a virtual machine (VM), attacks on package managers, data-based attacks, log file attacks, and Java application attacks.
\end{itemize}

\subsubsection{Network and Orchestration }

Container networks facilitate communication with external entities such as users or remote container systems and establish communication channels while managing isolation \cite{casalicchio2019container}. Orchestration, on the other hand, encompasses the coordination and management of container operations and services. Orchestration tools automate network configuration and traffic management \cite{casalicchio2019container}. Our findings identified fifteen risks and vulnerabilities associated with container networks and orchestration, as follows: 

\begin{itemize}
    \item \textit{Network risks and vulnerabilities}: They are insecure Sockets Layer
(SSL) strip, network resource authorization flaw, privileged network, vulnerable network interface (CNI) plugins, inefficient container network Configuration, leaking network policies, unfiltered traffic, misconfigured HostIPC, unverified packets in network path, ARP spoofing, MAC flooding attacks, escalation attacks, and MongoDB listening on the default port.
    \item \textit{Orchestration risks and vulnerabilities}: They include privileged access in Kubernetes, Uncontrolled Kubernetes Pods, and unreachable addresses in the Kubernetes overlay network.
\end{itemize}

%Table \ref{tab:Risks in Containers network and orchestration} presents container network and orchestration risks and vulnerabilities. 

\subsubsection{Runtime}

 Container runtime is a tool used to manage and execute containers to deliver the service in production \cite{ibrahim2019attack}. Our findings identified seven risks and vulnerabilities associated with the container runtime. According to the causes of the risks and vulnerabilities, as follows:
 %Table \ref{tab:Risks in ContainersRuntime} presents the risks and vulnerabilities in container runtime. They are categorized according to the causes of the risks and vulnerabilities, as follows:
 
\begin{itemize}
    \item \textit{Configuration risks and vulnerabilities}: they are the development-production environment and container runtime configuration.
    \item \textit{Development and production risks and vulnerabilities} they insecure code in the pipeline, SQL injection attacks, upgraded container images in production, vulnerable transmission control protocol (TCP)  timestamps, and deactivated user authentication.
\end{itemize}

\subsubsection{General container environment security }
In the general container environment, we collect the risks and vulnerabilities that do not belong to all the previous categories but still affect the container security. Our findings identified five general risks and vulnerabilities associated with container systems, which are grouped according to their inherent causes.

%Table \ref{tab:General container environment Risks and vulnerabilities} presents the general container risks and vulnerabilities grouped according to their inherent causes.

 \begin{itemize}
     \item \textit{Container security management}: it includes deficient permission management and vulnerable data analytical frameworks.
     \item \textit{External attacks to the container system} include ransomware and malicious third-party inputs.
     \item \textit{Power attacks}: it includes electrical workload threats to container systems.
 \end{itemize}

\subsection{Security Practices in Software Containers}
\label{What are the security practices and mitigation techniques to overcome risks and vulnerabilities in container systems?}

This section (addressing RQ2) explores security practices for improving container security. Security practices are activities that must be applied regularly to ensure the security of container systems. We searched the literature for such activities, which are grouped into vulnerability management practices, container security measures and considerations, security practices for isolation, image inspection, orchestration, and production security practices.
All the security practices discussed in this section do not follow the life-cycle taxonomy because they discuss the practices affecting the overall container system as one entity.

\textbf{Vulnerability Management Practices:} Vulnerability management is one of the main concerns in container security. It is concerned with detecting and mitigating vulnerabilities. Literature recommended continuous monitoring [P4] and continuous analysis of Docker Hub images to address the newly identified vulnerabilities and understand how to manage them  [P60]. Additionally, using Linux features to improve the container security and privilege division significantly decreases misconfiguration vulnerabilities [P91], [P6].
  
\textbf{Container Security Measures and Considerations:} Containers demand specific security considerations to mitigate potential risks. For example, the choice of the base image source and the inspection procedures should be applied to ensure that the image is safe or not high-risk [P72], [P83]. In addition, utilizing automated tools to enhance container security is a key practice to maintain a secure container environment at runtime and avoid human mistakes [P74], [P110]. Moreover, there are security measures that need to be highlighted to ensure system security, like determining the optimal timing for software updates [P1], inspecting the security of inherited packages [P4], and developing strategies to mitigate potential cloud attacks [P8], [P13].

\textbf{Security Practices for Isolation:} Applying isolation practices can significantly enhance container integrity. Such techniques involve configuring the functionalities necessary for SGX-based container orchestration and extending filesystem isolation to provide additional security layers [P25], [P87], [P127]. Another essential practice is properly configuring namespaces, which facilitates the application of autonomous security controls within containers and aids in detecting container escape attempts [P93], [P100], [P126].

\textbf{Image Inspection:} Scanning the container image is essential during container image execution \cite{azab2016software}. It is important for a bug-free container system, particularly if the container is set on a virtual machine [P33], [P2]. Also, rigorous inspection and management of image components, SW packages, codes, and all other dependencies added to the image help to secure the intra-container [P125]. Other practices that can improve image security are verifying the authenticity and integrity of Docker base images, choosing official images [P1], removing unused functions, and managing the different versions of the image [P89], [P7], [P105], [P72].

\textbf{ Application of Security policies:} It includes recommendations related to orchestration practices such as applying authentication and authorization rules to prevent unauthorized access, implementing security policies for pods and networks, scanning continuous delivery components for vulnerabilities, enabling logging and monitoring for clusters and networks, restricting access to internal databases, securing shared namespace sources, regularly updating clusters, and setting CPU and memory quotas [P56], [P86], [P15].

\textbf{Production Security Practices:} It includes implementing secure deployment practices for containers to enhance system security and fortify the continuous integration workflow [P46], [P25]. Also, utilizing control groups helps to improve resilience against resource exhaustion and DDoS attacks [P18], [P119]. Continuous runtime monitoring for security risks [P100], [P124] and employing system call monitoring techniques like the Bag of System Calls (BoSC) helps to detect abnormal behaviour [P113], [P52], [P58].

\subsection{ Tools used to improve container security}

This section (addressing RQ3) seeks to comprehensively examine the current landscape of tools designed for scanning and detecting various vulnerabilities within software containers. In addressing this research question, we meticulously scrutinize various tools documented in the literature by diverse researchers. Our data extraction process involves distilling pertinent information from state-of-the-art articles, encompassing details such as the phase in which the tool operates and the tool's purpose. 

Upon examination of Table \ref{tab:security Tools and Artifacts}, it is evident that we have identified 23 tools operating at varying levels of granularity to scrutinize and identify security issues in software containers. These tools are further classified into container life-cycle taxonomy as follows:
 \begin{itemize}
     \item Image tools that focus on identifying potential vulnerabilities and malicious packages in the image. It includes Clair, Microscanner, Notary, Vuls, Docker Scan, JFrog Xray, Banyanops Collector, Dagda, Grype and ConPan.
     \item Host Tools mainly focuses on vulnerability detection in containers and provides protection to application data. It includes AppArmor, Anchor, and CHAOS.
     \item Intra-Container tools that focus on detecting vulnerabilities in container dependencies and application isolation. It includes Trivy, Overshadow, Dagda, Grype, Snyk and Calico.
     \item Network tools that focus on securing the network and cloud connectivity, and detecting web server vulnerabilities. It includes Cilium, Nikto, Calico and AppArmor. 
     \item Runtime tools that focus on detecting abnormal system behaviour. It includes SPACE, Sysdig, Stools, and SCONE.
 \end{itemize}
Some tools can be used to secure multi-phases, like Dagda and Grype, used for both intra-container and image; Calico, used for container and network; and AppArmor used for network and host.

\begin{table*} [htp]
%\caption{Mitigation Techniques for Risks and Vulnerabilities}
%\label{tab:Mitigation techniques for security risks and vulnerabilities}
\caption{Security Tools and Their Purpose in Container Systems}
%Security Tools and Relevant Security Purposes

\label{tab:security Tools and Artifacts}
 \centering
\setlength{\tabcolsep}{4pt}
\begin{tabular}{|p{100pt}|p{100pt}|p{250pt}|}\hline
\textbf{Category} &\textbf{ Tool} &\textbf{ Purpose}  \\
\hline
Image  & Clair [P28] [P17] [P73] [P118] & A monitoring tool used to perform vulnerabilities static analysis for containers  \\
%\hline
  & Microscanner [P17] & A scanning tool for  image  package vulnerabilities \\
%\hline
  & Notary [P17] [P108] & A trusted tool provide mechanisms to sign and verify the image content  \\
%\hline
  & Vuls [P1] &  Linux vulnerability scanner used to detect vulnerabilities.  \\
%\hline
  & Docker scan [P69]&  A security scanner in Docker containers, used for image local scanning- This tool was enhanced and released under a new name Docker Scout - \\
%\hline
  & JFrog Xray [P73]  & Scan image for vulnerabilities and compliance issues [P73] \\
%\hline
   & Banyanops Collector [P108]  & Static analysis of container images \\
%\hline
  & ConPan [P24]  & Package vulnerability detection  \\
%\hline
% & AppArmor [P17]  & Prevent access to filesystem and network.   \\
Host  & Anchor [P17] & It provide encryption for the host files \\
%\hline
  & CHAOS [P76]  & Prevent the OSs from reading or tampering with application’s data \\
Intra-Container  & Trivy [P69] [P73]  & Detect vulnerabilities in image dependencies \\
%\hline
  & Overshadow [P76]   & Isolating application in different execution environments \\
  & Snyk [P121]   & Scan the dependencies in container package \\
Network  & Cilium [P17] [P108]
  & Securing network  and cloud connectivity \\
  & Nikto [P45]   & Web scanner for common web server vulnerabilities  \\

Runtime & SPACE [P31]  & monitoring too for container systems to detect abnormal behavior \\
%\hline
  & Sysdig [P69] [P108] [P21] & Monitor application in runtime and provide troubleshooting materials \\
%\hline
  & Stools [P1]   & Acting as a single layer Docker image to circumvent the Docker requirements in the Clair API1   \\
%\hline
  & SCONE [P76]  & Runs a container instance in a trusted execution environment based on Intel SGX \\

Intra-Container/Image  & Dagda [P69] [P108] 
  & Perform static analysis of known vulnerabilities and threats \\
%\hline 
 & Grype [P69] [P118] & Detect vulnerabilities in images and filesystems \\

Intra-Container/Network  & Calico [P17]  & Virtual networking security detection and policy enforcement  \\
Network/Host   & AppArmor [P17]  & Linux security module to prevent access to the filesystem and network  \\
\hline
\end{tabular}
%\label{tab1}
\end{table*}

\subsection{Publication Trends}
\label{When were the articles published?}

\begin{figure}
    \centering
    \includegraphics[width=1\linewidth]{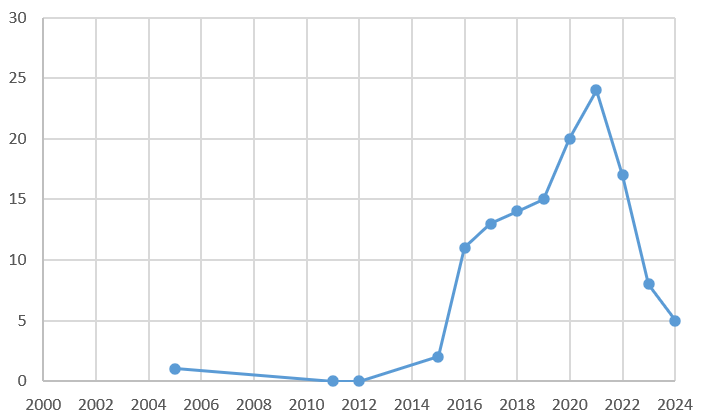}
   \caption{Number of Publications Over the Years}
    \label{fig:trend}
\end{figure}

Figure \ref{fig:trend} illustrates the publication trend concerning container security research from 2005 to 2024, addressing RQ4. Notably, there was a notable shortage of interest in container security research during the early period from 2005 to 2015, with only one published article between 2005 and 2014, reflecting a nine-year gap in publications addressing this critical topic. However, beginning in 2016, we can observe a gradual increase in the number of publications, with the highest number of publications (24) in 2021. 

The publication trend highlights research on software container security's growing relevance and significance over the past decade. The subsequent rise synchronizes with the increased adoption of container technologies after the release of the Docker platform in 2013. The trend also highlights the growing recognition of container risks and vulnerabilities and their implications in software engineering research literature.

\subsection{Publication Venues}

%Table \ref{tab:publication Venues on Container Security} summarizes the top ten publication venues, including conferences and journals, that have published more than one primary study.

In addressing RQ5, our analysis revealed that conferences are the predominant venue for publishing research on container security, with 98 primary studies published in conference proceedings, compared to only 31 studies published in academic journals. The top conferences are the \textit{IEEE International Conference on Software Analysis, Evolution and Reengineering (SANER)}, with six primary studies, and \textit{The IEEE International Conference on Cyber Security and Cloud Computing (CSCloud)}, with five primary studies. The entire list of publication outlets can be accessed here: \url{https://zenodo.org/records/15813355}.

These insights may help to identify the most common publication avenues for research focused on software container risks and vulnerabilities and to find the most relevant and impactful studies. Moreover, it highlights the diverse research context on software container systems concerning multiple domains and niches (e.g., cybersecurity, cloud computing, and hardware and infrastructures). 
\begin{comment}

\begin{table} [htp]
 \caption{ Publication Outlets}  
\label{tab:publication Venues on Container Security} 
\setlength{\tabcolsep}{2pt}
\begin{tabular}{|p{135pt}|p{50pt}|p{30pt}|}
\hline
Venue & Type & Number \\
\hline
 IEEE International Conference on Software Analysis, Evolution and Reengineering (SANER) & Conference  & 6 \\
IEEE International Conference on Cyber Security and Cloud Computing (CSCloud) &  Conference & 5 \\
IEEE Access & Journal &4 \\
International Symposium on Research in Attacks, Intrusions and Defenses (RAID) & Conference & 3 \\
Computers \& Security & Journal &3\\ 
International Conference on Utility and Cloud Computing & Conference& 2\\
Annual ACM Symposium on Applied Computing &Conference & 2\\
Annual Computer Security Applications Conference & Conference & 2\\
ACM SIGSAC Conference on Cloud Computing Security Workshop & Conference & 2 \\
International Conference on Cloud Computing (CLOUD) & Conference & 2 \\

  \hline

\end{tabular}
%\label{tab1}
\end{table}
\end{comment}

\subsection{Research Methodologies}

Empirical research is the approach to collecting and analyzing primary data to produce novel knowledge. It consists of methodologies such as surveys, interviews, and experiments. Meanwhile, non-empirical research refers to articles that do not analyze primary data and consist not only of opinion articles but also analytical, conceptual, and purely theoretical research. 

Upon addressing RQ 6, we identified nine empirical research articles that explicitly reported the methodology used: eight applied experiments, while one employed a quantitative survey. Although many other studies included evaluations of proposed tools or techniques, they were not considered empirical research because they did not include empirical evidence.

The remaining articles did not clearly report their methodological approach. Therefore, we grouped these articles based on their stated research focus, as follows:

\begin{itemize}
    \item Proposing/developing novel security frameworks that propose a structured approach to mitigate certain security issues (72).
    \item (Overall) security enhancement techniques (28).
    \item Explains risks/vulnerability analysis process (16).
    \item Employing threat modelling techniques (4).
\end{itemize}

 Mapping the employed methodologies trends in container risk and vulnerabilities reveals the low number of empirical studies, indicating a need to validate the current knowledge on container risks and vulnerabilities. Also, the studies proposing new security frameworks indicate the developing need for practical solutions to current security challenges.

\section{Discussion}
\label{Discussion}

The analysis of primary studies on container security issues shows that researchers view container systems as separate and loosely coupled entities rather than a chain of interconnected phases, where each phase significantly impacts the following phase. Most of the primary studies emphasize risks and vulnerabilities associated with intra-container, networks and orchestration but do not give equal importance to the risks and vulnerabilities associated with the image, host, and runtime phases of the container life-cycle. It is important to pay attention to image, host, and runtime risks and vulnerabilities because they may threaten the whole container system. For example, a vulnerable image can cause an escalated container privilege; the misconfigured host can expose the network ports and facilitate attacks, and malicious SQL injection in the runtime can cause database modification. Consequently, future research should focus more on the non-investigated phases of the container life-cycle, including the image, host and runtime risks to enhance the security features.

As seen in Section \ref{RQ: What are container security risks and vulnerabilities?}, most of the causes for risks and vulnerabilities stem from misconfiguration issues or configuration flaws. In a nutshell, overlooking these configuration flaws can create opportunities for multiple attacks and threats to a container life-cycle's phases, eventually leading to a catastrophic failure of the entire system during runtime. Misconfiguration issues or flaws, in essence, highlight the importance of human decision-making concerning the development of container systems. This underscores the importance and necessity of security training for practitioners to develop highly secure container applications.

The findings on security practices provide valuable insights into the current container security management. It highlights how security practices used to improve the security of the overall container system and how they can be effectively developed into proactive safety procedures in future. The findings underline the idea that security should be embedded in the early phases of container development. On the other hand, there were some observations on how security practices were discussed in SE literature. For example, articles often fail to explain how to practically execute processes and practices to improve container security. Instead, the articles only list the security practices without explaining how the process is executed. Similarly, the articles concerning image scanning processes do not clearly describe how the scanning process is executed; instead, these articles only suggest conducting ìmage scanning' to enhance the system's security. Subsequently, future research on software containers must also focus more on improving and describing the processes themselves and not just prescribing the practices. A strong collaboration between academia and industry could bridge this research gap. %and emphasise utilizing official trusted images. In addition, it fosters continuous scanning processes for images and misconfigured access rights to mitigate unauthorized modifications.

%The findings on runtime monitoring provide practical directions on how security is managed in production. Monitoring logs provides oversight about potential risks and vulnerabilities that need a rapid response in deployment. Moreover,  integrating automated dynamic scanning in CI/CD pipelines would alert developers to potential risks. In addition, implementing isolation mechanisms and maintaining user rights in production are highly recommended to ensure unauthorized access. Overall, we could make the container systems more secure by incorporating mitigation techniques for risks and vulnerabilities and implementing security enhancement practices during the development and deployment processes.  

%Although the findings on security practices are valuable, there are concerns about how security practices are discussed in the literature. For example, articles often fail to explain how to practically execute processes and practices to improve container security. Instead, the articles only list the security practices without explaining how the process is actually executed. Similarly, the articles concerning image scanning processes do not clearly describe how the scanning process is executed; instead, these articles only suggest conducting ìmage scanning' to enhance the system's security. Subsequently, future research on software containers must also focus more on improving and describing the processes themselves and not just prescribing the practices. A strong collaboration between academia and industry could bridge this research gap.

We also noticed that articles use different terms to describe the same tool function. For example, the terms ``vulnerability scanners'' and ``vulnerability detection'' are used to describe the tools used to discover vulnerabilities in container systems. Using different descriptions for the tools was confusing and raised questions about the tools' function in containers. 

The identified security tools included open-source and commercial options. Open-source tools like Clair, Microscanner, Trivy, Calico, Dagda, grype, Notary, Vuls, DockerScan, Banyanops Collector, ConPan, Anchore, Snyk, Cilium, Nikto, and AppArmor are free and open-source tools. These tools are also customizable to meet diverse companies' security requirements. However, free tools are effective in practice, and they do not always fulfil the required security needs. Therefore, some commercial tools are available to provide advanced features, such as JFrog Xray \footnote{https://jfrog.com/}, Sysdig \footnote{https://sysdig.com/}, and SCONE\footnote{https://azuremarketplace.microsoft.com/ens/marketplace/apps/ \\ scontainug1595751515785.scone?tab=overview}. These tools support practices such as static analysis, runtime monitoring, network security, and vulnerability scanning. Companies tailor a mix of these tools to satisfy container security requirements 

Our study revealed that existing literature discusses the risks and vulnerabilities associated with different phases of container systems. However, these discussions lack comprehensive mitigation techniques for some of the identified risks and vulnerabilities. This observation underscores the need for future research to investigate more into developing robust container mitigation techniques. Furthermore, our analysis of the tools covering the container life-cycle showed an extensive focus on image vulnerability checks compared to other phases in the life-cycle. This finding raises concerns about the security of the overlooked phases and calls for future research to prioritize developing new tools to secure all container phases.

We also observed from our findings that most primary studies (76\%) are published in international conferences, and (24\%) of articles are published in journals. This trend could indicate the accelerating development in container security, and researchers target conferences because they provide a faster publication channel.
 
\subsection{Implications for Future SE Research}
The research reviewed in this article does not support any concrete and practical techniques to overcome the risks and vulnerabilities issues in software containers. Better recommendations will require the following investigations:

\begin{enumerate}

\item \textit{Practitioners' perspective on managing container security:} This future research will comprehensively examine the perspectives and strategies of practitioners in managing container security. By exploring real-world challenges and practices, this study seeks to identify gaps and develop evidence-based management frameworks that enhance container security protocols and operational policies. A detailed understanding of these challenges will guide the development of concrete, practical security approaches that align with industry needs.

\item \textit{Advancing testing methodologies for anomaly detection in software containers:} This future research will focus on assessing and enhancing current testing methodologies for detecting anomalies within containerized environments. The goal is to identify limitations in existing approaches and propose advanced testing frameworks that improve the detection and mitigation of potential risks. Enhanced anomaly detection mechanisms will contribute to a more resilient security measure in container systems.

\item \textit{Developing a framework for prioritizing vulnerabilities in container systems:} The primary goal for this future research is to create a systematic framework for assessing and prioritizing vulnerabilities and risks within container systems. Emphasis will be placed on the early identification and mitigation of high-risk vulnerabilities during the implementation phase, thus minimizing exploitation risks and reinforcing overall system security. This approach will ensure that critical vulnerabilities are addressed proactively, supporting the reliability of containerized applications. 
  
\item \textit{Evaluating and enhancing the current security practices in container systems:} This future research will conduct an in-depth analysis of existing security practices utilized in containerized systems, assessing their effectiveness and potential for improvement. By identifying successful and suboptimal practices, the study will propose refinements and validate them through targeted case studies and real-world use cases. This investigation shall advance current practices and inform future advancements in container security strategies.

\end{enumerate}

\subsection{Implication for SE Practitioners}
 The research findings contribute to SE practice as follows:
 \begin{enumerate}
     \item \textit{Improve and enhance SE practitioner knowledge on container security:} This study would help practitioners understand potential risks, vulnerabilities' root causes, implications, risk mitigation strategies, and improve system security using practices and tools. It will help practitioners avoid the pitfalls and improve security decisions during implementation.
     
     \item \textit{Security issues within interlinked software container life-cycle:} The study shows that the phases of container development are interlinked. Security issues in one phase can impact that phase and extend to other phases. Practitioners need to consider the causes of security issues and proper configurations of container systems to achieve robustness across all phases of the container life-cycle. 
     
    \item \textit{Explore various container security tools:}  Practitioners can better understand how to use the various container security tools by focusing on their functionalities, appropriateness, and the phases of container development where they could be used. This will help practitioners choose a suitable tool and customize it according to the project's needs.
    
 \end{enumerate}

\subsection{ Limitations and Threats to Validity}
We reflect on the quality of systematic mapping studies by following the suggestions made in \cite{montgomery2022empirical} and \cite{easterbrook2008selecting}, as  follows:

\textit{Oversimplification of the content:} Reporting technical security issues in container systems through a qualitative study required simplifying technical information. This may have limited the depth of detail, particularly regarding mitigation techniques and their implementation.

%Exploring the risks and vulnerabilities of technical aspects like container systems security and reporting the findings qualitatively introduces a situation where complex ideas and technical information were required to be presented in an oversimplified manner. This might negatively impact the details of our findings, specifically concerning the mitigation techniques and their implementation. This phenomenon might have also resulted in the loss of other relevant articles from our analysis.

\textit{Internal validity:} To ensure internal validity, we followed the guidelines proposed by \cite{kitchenham2010systematic, petersen2015guidelines, ralph2020acm} and reported all steps (Section~\ref{studydesign}). Screening criteria were refined collaboratively, and multiple pilots were conducted to support consistency in study selection and data extraction.
%The internal validity refers to the extent to which we followed the guidelines to conduct the study. To maintain the internal validity, we followed the guidelines as proposed by \cite{kitchenham2010systematic}, \cite{petersen2015guidelines}, and \cite{ralph2020acm} by reporting each step in the research process as described in Section \ref{studydesign}. The authors' team revised the screening criteria to maintain consistency in the primary study selection process. We also conducted multiple pilots to ensure consistency and coverage during the study selection process and while extracting the relevant data.
 
\textit{External validity:} To maintain external validity, we selected widely used SE databases (IEEE Xplore, ACM Digital Library, Scopus, Web of Science, ScienceDirect) as suggested by \cite{dyba2007applying} to broaden the search coverage. Moreover, we conducted three pilots to identify the most appropriate search string to increase the coverage of our results. While comprehensive, the possibility remains that some relevant studies were missed due to limitations in string construction or database scope (see Section~\ref{Literature search}).
%External validity was maintained by choosing the most commonly searched databases in SE studies (as suggested by \cite{dyba2007applying}), including IEEE Xplore, ACM Digital Library, Scopus, Web of Science, and ScienceDirect. This helped to broaden the search coverage. Moreover, we conducted three pilots to identify the most appropriate search string to increase the coverage of our results. This was done by comparing the retrieved studies against a known list of articles (i.e., relevant to our research). Additionally, we also compared the outcomes of different trial searches to verify whether they encompassed these known studies and whether any new,
%significant primary studies emerged. However, it is plausible that certain relevant studies may not have been included due to incomplete search strings or the omission of certain essential databases. Search strings, specific filters applied in databases, and results from our search are reported in the section describing our study design (refer to Section \ref{Literature search}). The entire author team also discussed
%the breadth of coverage offered by the various trial searches. Finally, we evaluated and considered the methods and findings of
%related previous secondary studies.

\textit{Construct validity:} To support construct validity, all authors discussed and defined key concepts, including the risks, vulnerabilities, security practices, tools, and container life-cycle, to maintain a consistent understanding of these concepts across the article and thereby avoid researcher bias. 
%The construct validity of this study focuses on interpreting its main concepts and terminology. To ensure construct validity, the author team discussed all the relevant data items that were to be extracted from the primary studies. We also define the key terms used in this study, including the risks, vulnerabilities, security practices, tools, and container life-cycle, to maintain a consistent understanding of these concepts across the article and thereby avoid researcher bias. Consequently, adjustments were made to the names, explanations of extraction questions, and extraction procedures to synchronize with the author team’s interpretations. Furthermore, low-confidence instances were identified and discussed between the first two authors until a consensus was reached. For example, we discussed and agreed on the discrepancies between security enhancement practices and risk mitigation techniques.

\section{Conclusion}
\label{Conclusion}
This study makes significant contributions to understanding the current software container security landscape in SE research by systematically mapping the risks and vulnerabilities present across various phases of the container life-cycle. We identified multiple risks and vulnerability issues in container security across a total of 129 relevant studies published between 2000 and 2024. The major takeaways of this study are as follows:

\begin{enumerate}

    \item Our study organizes and presents a comprehensive and novel taxonomy categorizing risks and vulnerabilities in software container systems, addressing a previously fragmented body of knowledge. This provides a valuable reference framework for researchers and practitioners.
    %, enabling more targeted and comprehensive studies in the field of container security.

    \item Furthermore, the findings highlight existing gaps, particularly the need for more research into critical phases of the container life-cycle, such as image, host, and runtime, which have been underrepresented compared to intra-container and network phases. This insight directs future research efforts toward under-investigated areas.
    %supporting balanced advancements toward research in container security.

    \item By analyzing the current landscape of container security in terms of the causes, effects and mitigation techniques of the risks and vulnerabilities, the study encourages further development of practical, empirically-informed security enhancement measures and enhances the practical application of security research. 
    %Consequently, the researchers are motivated to focus on refining and empirically validating strategies that go beyond theoretical prescriptions, thereby enhancing the practical application of security research.

    \item The findings’ emphasis on the interconnected nature of container life-cycle phases also highlights the importance of developing holistic security strategies. This perspective encourages the research community to approach container security as a cohesive system.
    %fostering advancements that strengthen all life-cycle phases rather than focusing narrowly on isolated issues.
    
\end{enumerate}

These contributions collectively guide the software engineering research community toward more integrated and practice-oriented investigations on software container security concerns, thereby bridging the gap between theoretical insights and real-world applications.

\section*{Acknowledgment}
This work was supported by Containers as the Quantum Leap in Software Development (QLeap) project, funded by Business Finland (BF) grant, number 3215/31/2022.

Copilot \footnote{https://copilot.microsoft.com/} is the AI tool used for editing and grammar enhancement.

%\textit{Containers as the Quantum Leap in Software Development (QLeap)} 
%3215/31/2022
%\bibliographystyle{IEEEtran}
%\bibliography{Ref}

\bibliographystyle{IEEEtran}
\bibliography{Refer}

\end{document}